\documentclass[a4paper,twocolumn]{revtex4}
\usepackage{graphicx}
\usepackage{float}

\begin{document}

\title{Spin density distribution in open-shell transition metal systems: A comparative post-Hartree-Fock, Density Functional 
Theory and quantum Monte Carlo study of the CuCl$_2$ molecule}

\author{Michel Caffarel}
\author{Emmanuel Giner}
\author{Anthony Scemama}
\author{Alejandro Ram\'irez-Sol\'is}
\thanks{On sabbatical leave from Facultad de Ciencias, Universidad Aut\'onoma del Estado de Morelos, M\'exico}
\affiliation{CNRS-Laboratoire de Chimie et Physique Quantiques, IRSAMC. Universit\'e Paul Sabatier, 118 route de Narbonne, 31062 Toulouse Cedex, France}

\begin{abstract}
We present a comparative study of the spatial distribution of the spin density (SD) of the ground state of CuCl$_2$ using Density Functional 
Theory (DFT), quantum Monte Carlo (QMC), and post-Hartree-Fock wavefunction theory (WFT). A number of studies have shown that an accurate 
description of the electronic structure of the lowest-lying states of this molecule is particularly challenging due to the interplay 
between the strong dynamical correlation effects in the $3d$ shell of the copper atom and the delocalization of the $3d$ hole over the chlorine atoms. 
It is shown here that {\it qualitatively} different results for SD are obtained from these various quantum-chemical approaches. 
At the DFT level, the spin density distribution is directly related to the amount of Hartree-Fock exchange introduced in hybrid functionals and,
therefore, it is not possible to draw conclusive results. At the QMC level, Fixed-node Diffusion Monte Carlo (FN-DMC) results for SD 
are strongly dependent on the nodal structure of the trial wavefunction employed. In the case of this open-shell system, 
the $3N$-dimensional nodes are mainly determined by the 3-dimensional nodes of the singly occupied molecular orbital (SOMO) and FN-DMC results 
are found to be strongly dependent on the type of one-particle model used for generating the SOMO (here, Hartree-Fock or Kohn-Sham 
with a particular amount of HF exchange). 
Regarding wavefunction approaches, HF and CASSCF lead to strongly localized spin density on the copper atom, in sharp contrast with DFT. 
To get a more reliable description and shed some light on the connections between the various theoretical descriptions, Full CI-type (FCI) 
calculations are performed. To make them feasible for this case a perturbatively selected CI approach generating multi-determinantal expansions of 
reasonable size and a small tractable basis set are employed. Although semi-quantitative, these near-FCI calculations allow to clarify how the spin density 
distribution evolves upon inclusion of dynamic correlation effects. A plausible scenario about the nature of the SD is proposed.
\end{abstract}

\maketitle
\section{Introduction}
In spite of much effort in the last 50 years, to devise a general electronic structure approach that is both computationally practical 
and accurate enough for all types of molecular systems is still a challenging task. Indeed, to provide a truly accurate account 
of the electronic structure of a molecule one must take into account in a {\it balanced} way several effects of different physical/chemical nature, 
a) electron-electron correlation effects (resulting from the $1/r_{12}$ interaction), b) exchange effects (Pauli principle), 
c) delocalization (kinetic effects) and, in some cases, d) quasi-degeneracy effects (quantum entanglement of almost degenerate 
low-energy wavefunction components). All the present-day methods deal with these aspects in different ways, sometimes with not so-clear distinctions 
between them ({\it e.g.} the mixture of exchange and non-dynamical correlation effects within Kohn-Sham formulation of Density Functional theory).
Here, in our study we shall consider the two most widely used electronic structure methods, namely 
Density Functional Theory (DFT) and molecular orbital-based or wavefunction theories (WFT) (post-Hartree-Fock approaches). 
We shall also consider quantum Monte Carlo (QMC) methods that are potentially very accurate but
are still methods of confidential use in quantum chemistry due to a number of practical/theoretical limitations.
Each type of method treats the various effects cited above in different ways with particular strengths and weaknesses.

Density functional theory (DFT) is nowadays the most popular and widely used theory for the description of 
electronic structure of atoms, molecules and condensed phases (solids and liquids). Its success stems 
mainly from the fact that it provides reasonable energetic and structural properties at a moderate
computational cost. One of the most fundamental tenets of DFT is that it expresses the general $N$-body interacting electronic system 
into a much more simple effective one-body system where electrons are immersed into a universal exchange-correlation potential embodying all the 
complexity of the $N$-body problem. Within this formalism 
all types of electronic  systems, regardless of the strength and characteristic length of the interactions involved 
({\it i.e.} covalent, ionic, hydrogen-bonded, metallic or van der Waals types) are expected to be treatable, at least in principle. 
However, many questions remain open in the DFT realm, mostly due to the necessity of approximating in a coherent way the unknown universal 
exact exchange-correlation functional. In the last 20 years DFT has made very important progress, mainly due to the improvements 
over Generalized Gradient Approximation (GGA-type) functionals: The meta-GGA functionals\cite{1,2,3} which use the 
kinetic energy densities ($\tau_{\uparrow},\tau_{\downarrow}$) and/or the Laplacian of the density as non-local variables, 
and hybrid adiabatic connection functionals\cite{Becke-JCP93} that empirically mix Hartree-Fock (HF) non-local 
exchange with GGA exchange. However, all the hybrid functionals contain several adjustable semi-empiric parameters. 
Some of the hybrid functionals have become extremely popular since they yield, in general, rather good energetic 
and structural results for main-element molecules. Nevertheless, fundamental problems still appear, for example the fact that many 
of these semi-empirically adjusted exchange-correlation (xc) functionals do not even provide the correct asymptotic exchange-correlation potential 
for systems for which the exact behavior is known.\cite{JP-LecNotes}

In the case of WFT the quantum chemical description passes through the construction of an explicit wavefunction with 
the need of accurately introducing static and dynamic electronic correlation effects. Ideally this can be achieved through 
the construction of the Full Configuration Interaction (FCI) wavefunction. However, since for most molecules the FCI solution is 
readily out of reach even with moderate basis sets (exponential increase of the Hilbert space), 
approximate solutions are needed and are achieved in practice by building increasingly complex 
wavefunctions following one of several approximations using either perturbation, truncated CI (CIS, CISD, CISDQ, CISDTQ,...) 
or coupled cluster (CCSD, CCSD(T), ..) techniques. Note hat, within the WFT framework, the question of whether or not the electronic 
state in question can be correctly described using single-reference methods also appears. In the negative case the application 
of the Complete Active Space SCF (CASSCF) method has become customary and the ensuing CASSCF wavefunction is used as zeroth-order 
reference for further treatment of the dynamic correlation effects, for instance, through the CASPT2 or the Averaged Coupled Pair 
Functional (ACPF) methods. At this point we emphasize that for the spectroscopic properties of transition metal molecules 
such as the CuCl$_2$ molecule treated here, critical accuracy problems might arise 
due to the quality of the atomic basis sets used or 
to the quality of the electronic correlation treatment. Unfortunately, the application of very accurate wavefunction methods is still 
restricted to molecules with a few atoms due to the rapid growth of the computational cost with the number of electrons 
(or the number of atomic basis functions). 

The third type of methods considered here are the so-called quantum Monte Carlo (QMC) approaches. QMC are statistical methods for 
solving the Schr\"odinger equation. They are very attractive since they are potentially exact methods 
(up to the statistical errors inherent to any Monte Carlo approach). Unfortunately, in practice we have to cope with the 
pathological fluctuations of the wavefunction sign and a so-called fixed-node approximation has been introduced to fix this problem.
This approximation can be viewed as solving the electronic Schr\"odinger equation but with a new additional constraint, namely, imposing the
solution to vanish wherever a known trial wavefunction given as input vanishes. In other words, the nodal hypersurface of the fixed-node 
wavefunction (nodes= $3N$-dimensional hypersurface where the wavefunction vanishes) are imposed to be identical to those of the approximate 
trial wavefunction. Numerical experience has shown that the fixed-node error is small according to the quantum chemistry standards
(typically, a small percentage of the correlation energy for total energies) but, unfortunately, still large enough to lead to potential difficulties 
when computing the small energy differences involved in quantitative chemistry. Stated differently, suitable cancellation of fixed-node errors 
are needed. At this point, it is worth emphasizing that the need of suitable cancellation of errors in energy differences is of course
not restricted to QMC and is, unfortunately, one of the most critical issues common to all computational chemistry approaches developed so far.
In practice, it has been observed that the nodal quality is directly related to 
the physical/chemical content of the trial wavefunction. In short, the better the trial wavefunction is,
the smaller the fixed-node error is. Let us emphasize that the need of having a rather good trial wavefunction to start a QMC 
calculation brings back some heuristics into the approach, a crucial point one has to be aware of. This aspect will be exemplified in
in the case of the CuCl$_2$ molecule treated in this work. 

As seen, for different reasons none of these state-of-the-art approaches are fully satisfactory to deal with all types of molecular problems.
Here, we propose to shed some light on their theoretical and practical relationships on a small molecule which is representative of 
a difficult molecular problem, namely, the ground state properties of the CuCl$_2$ molecule. As shown in previous studies, 
even the determination of the nature of the ground state and the proper energetic ordering of 
the low-lying states of this molecule turns out to be particularly difficult. This is mainly due to a subtle interplay between the delocalization of 
the Cu(3d) hole on the molecular axis and the dynamic correlation effects. Here, in order to investigate such relationships we focus on 
the spatial distribution of the spin-density of the ground-state along the molecular axis, which is the main physical quantity associated 
to the relative stability of the lowest electronic states in CuCl$_2$. More precisely, we consider the difference of $\alpha$ and $\beta$ spin densities 
integrated within the plane perpendicular to the molecular axis (actually, a parallelepiped of small thickness). Our working definition is
\begin{equation}
\Delta \rho(z) = \int_{z-\epsilon/2}^{z +\epsilon/2} dz \int\int dxdy [\rho_{\alpha}({\bf r}) - \rho_{\beta}({\bf r})],
\label{deltarho}
\end{equation} 
where $z$ is the coordinate along the molecular axis of the linear centro-symmetric molecule, the copper atom being at the origin, and $\epsilon$ 
is a small positive parameter (here, chosen equal to 0.1 a.u.) corresponding to the thickness of the parallelepiped. For simplicity this 
partially integrated difference of $\alpha$ and $\beta$ densities will be shortly referred to in the present work as the spin density (SD) distribution.
Although the DFT approaches lead to a slightly different optimized geometry ({\it ca.} 3.92-3.97 a.u.), in what follows we shall use 
the equilibrium centro-symmetric geometry fixed at a Cu-Cl distance of 3.9 a.u., closer to the experimental value of 3.85 a.u. for the ground state.

The contents of the paper are as follows. In Sec. \ref{cucl2} we summarize what is known about the nature of the low-lying electronic 
states of CuCl$_2$. Sections \ref{dftrho},\ref{wftrho},\ref{qmcrho}, and \ref{cirho} present the results obtained for the spatial distribution of the 
spin density using DFT, WFT, QMC and near-Full CI, respectively. Finally, in Sec. \ref{discussion} a detailed summary 
and discussion of the results obtained is presented. In particular, a plausible description of the nature of the ground state is proposed.

\section{What is known about CuCl$_2$}
\label{cucl2}
The quantitative description of the electronic structure of metal-containing systems is known to be a
rather delicate problem.
Regarding DFT studies, we can mention three articles reporting extensive tests of a series of functionals on transition metal (TM) containing molecules.
Schultz {\it et al.}\cite{Schultz} propose a data set of bond lengths for 8 selected 
TM dimers, their atomization energies and use these for testing DFT, while Furche and Perdew\cite{FP-JCP06}
also investigate the performance of contemporary semi-local (GGA, meta) and hybrid density functionals for 
bond energetics and structures of 3d transition-metal compounds. In this direction, one of us reported\cite{ARS-JCP07} 
a systematic study on the performance of local, semi-local and non-local xc functionals for the lowest 
singlet-triplet transition in AgI, showing that a rather unpredictable and thus, unreliable, 
performance of the various types of functionals appears for this singlet-triplet transition.
Although the CuCl$_2$ molecule was not included in these studies, its spectroscopy presents a 
particularly difficult case for {\it ab initio} and DFT methods, since important correlation effects 
(arising mainly in the 3d shell of copper) are strongly coupled to charge transfer effects via the 3p orbitals 
of the Cl ligands. Fortunately, the low-lying transitions 
are experimentally quite well known\cite{Crozet,Bondibey} and two extensive benchmark variational 
multireference Averaged Coupled Pair Functional(CASSCF+ACPF) studies on the spectroscopy of CuCl$_2$ exist; 
there the attention was focused on the nature of the three lowest electronic states\cite{RamDau-JCP04,RamDau-JCP05} 
that give rise to the four observed transitions. 
The first three ligand field (LF) states are thought to arise from d-d transitions on the copper ion and they can be described by 
a different orientation ($\sigma$, $\pi$ or $\delta$) of the 
singly occupied HOMO, in principle, the localized Cu(3d) hole. 
So, at this point one might ask why is this such a complicated problem?
In order to understand the complexity in the spectroscopic description involving the five lowest ligand-field (LF) and charge-transfer (CT) 
states note that, at the doubly ionic limit, CuCl$_2$ is described by the Cl$^-$Cu$^{2+}$(3d$^9$)Cl$^-$ structure, while in the covalent ClCuCl 
description, the copper atom which is promoted to the 3d$^9$4s$^2$ excited state undergoes 4s-4p hybridization and can establish covalent 
bonds with both Cl atoms. An intermediate situation arises when one considers the resonant Cl$^-$Cu$^+$(3d$^9$4s$^1$)Cl and ClCu$^+$(3d$^9$4s$^1$)Cl$^-$
ionic structures. Near the equilibrium geometry, the exact electronic structure for all states is a mixture of these three valence situations. 
The first three LF states ($^2\Sigma_g^+$, $^2\Pi_g$, $^2\Delta_g$) correspond to d-d transitions on the copper ion and it is generally thought 
that they can be described by the $\sigma$, $\pi$ or $\delta$ orientations of the singly occupied Cu(3d) orbital. It is known that a correct 
description of electronic structures, and even more with such close lying states, must include a correct description of correlation effects 
especially important for the d shell, but also must allow for large repolarization differential effects between localized d-d states and 
charge transfer states. We stress that single-reference methods like Coupled Pair Functional(CPF) and CCSD(T) can be used here, 
since the HF wavefunctions are excellent zeroth-order approximations for the lowest electronic states of CuCl$_2$ \cite{BR-JCP89}. From the 
DFT perspective, this feature is also very convenient, since standard Kohn-Sham based methods are, in principle,  well adapted to describe 
transitions where only a change in the orientation of the 3d-hole in the central metal atom is involved. 
We stress that the $^2\Pi_g \rightarrow ^2\Sigma_g^+$ transition in CuCl$_2$ represents a most difficult problem from the 
quantum theoretical point of view, since it has been predicted to range from -2495 to 6930~cm$^{-1}$.\cite{RPVD-JCP05}

On the other hand, since the lowest five electronic states (all doublets) of CuCl$_2$ belong to
different spatial symmetries, it has been possible to study these states through the $\Delta$SCF approach in the 
DFT framework\cite{Salahub,Deeth,RPVD-JCP05}. In this approximation each state, which is the ground state for a 
given spin and space symmetry, is optimized independently. Therefore, the main goal of\cite{RPVD-JCP05} was to perform a coherent 
assessment of DFT results with benchmark {\it ab initio} calculations, eliminating the discrepancies found in previous studies 
so that the comparison was restricted to the basic ideas of DFT. The DFT calculations were done with the same relativistic effective core potentials (RECPs)
 and optimized extended basis sets used in Refs.\cite{RamDau-JCP04} and \cite{RamDau-JCP05}. 
Table \ref{tab1} presents some selected DFT (LDA,GGA, hybrid and meta) and the {\it ab initio} $^2\Pi_g \rightarrow ^2\Sigma_g^+$
transition energies along with the corresponding spin densities on the central Cu atom.
\begin{table}
\centering
\begin{tabular}{|c|c|c|c|}
\hline
\hline
Method & $^2\Pi_g \rightarrow ^2\Sigma_g^+$ & $^2\Pi_g$ SD & $^2\Sigma_g^+$ SD\\
\hline
\hline
LDA(S+VWN5) & 6539 & 0.316 &\\
\hline
BLYP  & 4802 & 0.429 & 1.04  \\
\hline
PBE96 & 4699 & 0.43 & 1.03  \\
\hline
HCTH407& 4345 & 0.420 &\\
\hline
OPTX-LYP & 3963 & &\\
\hline
TPSS  &  4065 & 0.406 & \\
\hline
M06-2X & 3251 & 0.648 & \\
\hline
B3LYP&  1703 & 0.57 & 1.07\\
\hline
B97-2&  1465 & 0.54 & 1.03\\
\hline
PBE0 &   756 & 0.64 & 1.08 \\
\hline
NR-SCF$^a$ &  -2495 & 0.962 &\\
\hline
CASSCF(21,14)& 6930 & 0.94 &  1.00 \\
\hline
CASSCF+ACPF& 232 & & \\
\hline
CASPT2   &   3861 & &\\
\hline
NR-SDCI$^a$ & -2116 & &\\
\hline
NR-SDCI+Q$^a$ &  -1856 & &\\
\hline
CCSD(T) &   859 & &\\
\hline
NR-Coupled Pair Functional$^a$ &  659 & &\\
\hline
Theor.$^b$ &  900 & &\\
\hline
Exp.$^c$ &  253, 303, 475& &\\
\hline
\hline
\end{tabular}
\caption{DFT, {\it ab initio} and experimental transition energies in wavenumbers, Mulliken spin densities (SD) on the Cu atom where available.\\
\noindent
 $^a$ Non-relativistic calculations from \cite{BR-JCP89}.\\
 $^b$ Theoretical spin-orbit deconvoluted value from \cite{RamDau-JCP04}.\\
 $^c$ Experimental fine-structure transition energies; see corresponding references in \cite{RPVD-JCP05}.\\
}
\label{tab1}
\end{table}
Note that, within the {\it ab initio} framework, the dynamic correlation effects that control the nature of the Cu 3d-hole 
in the ground state are extremely difficult to obtain correctly since the SCF, the SDCI, and even the usually very accurate 
SDCI+Q (with Davidson's approximate size-consistent correction) schemes, all wrongly lead to a $^2\Sigma_g^+$  ground state. 
Only more sophisticated size-consistent {\it ab initio} methods 
like the Coupled Pair Functional (CPF), CCSD(T) or CASSCF+ACPF are able to correctly predict a $^2\Pi_g$ ground state, lying 659, 859 and 232~cm$^{-1}$
(respectively) below the $^2\Sigma_g^+$ one without spin-orbit (SO) effects. Note that, at the purely electronic level, 
transition energies must be compared with the theoretical SO-deperturbed value, estimated to be 900~cm$^{-1}$.\cite{RamDau-JCP04}  
We also stress that Bauschlicher and Roos\cite{BR-JCP89} showed that the Darwin and mass-velocity relativistic effects cancel out 
nicely for the spectroscopy of this molecule and, for this reason, they were able to use all-electron non-relativistic (NR) calculations ; 
in what follows we shall also make use of this fact. From the DFT perspective, most functionals (GGA, hybrid and even meta-ones) 
like HCTH407, BLYP, PBE96, OptX-LYP, TPSS and M06-2X largely overestimate this transition, all yielding values above 3200~cm$^{-1}$. 
Note that up to date, it is impossible to decide {\it a priori} which functional is to be used and which one can be trusted to 
yield reliable transition energies for an arbitrary metallic molecule. 
The delicate issue of the parametrization of most exchange-correlation functionals without the inclusion of transition metal 
containing systems has been discussed\cite{RPVD-JCP05} in this context. It is somewhat ironic that much less expensive and sophisticated 
descriptions such as those given by the PBE0 (750~cm$^{-1}$) and the B97-2 (1400~cm$^{-1}$) functionals yield better approximations to 
this transition energy than the very computational demanding benchmark CASSCF+ACPF one at 232~cm$^{-1}$. 
So the natural question arises: Are these hybrid PBE0 and B97-2  densities 
correctly describing each electronic state, therefore providing truly accurate total energies, 
or is this energy difference hiding some cancellation of errors associated with physically 
relevant quantities, such as the spatial distribution of charge and spin densities?    

\begin{figure}
\includegraphics[width=\columnwidth]{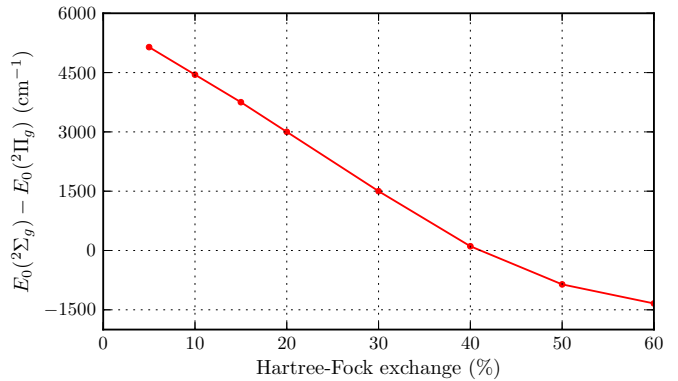}
\caption{ $^2\Pi_g \rightarrow ^2\Sigma_g^+$ transition energy (wavenumbers) with B3LYP as a function of the HF exchange percentage employed; positive values correspond to a $^2\Pi_g$ ground state, in agreement with experiment.}
\label{fig1}
\end{figure}

Although the various results presented in Table \ref{tab1} may appear rather diverse, a clear 
trend can be perceived in the DFT subset, namely the role of the amount of 
HF exchange in the relative stability of both states. The larger the percentage of HF exchange 
in the functional the more stable the $^2\Sigma_g^+$ state becomes.
In the extreme case of SCF-HF, we find the largest (negative) transition energy with $^2\Sigma_g^+$ state as the ground-state. 
In the opposite case where no HF exchange is included (BLYP, for instance), the $^2\Pi_g^+$ state becomes the ground state with a large (positive)
transition energy. In between, one can see that for functionals having a fraction of HF exchange, this transition energy is still positive but smaller.
To illustrate quantitatively this idea we present in Figure \ref{fig1}, for the B3LYP functional with variable HF exchange, 
the evolution of the transition energy {\it vs.} the HF exchange percentage.\\
This figure illustrates in a particularly striking way the high level of arbitrariness present when using hybrid functionals, such as B3LYP, for this system.
Clearly, there is no rational way to decide which is the ``right'' amount of HF non-local exchange to be used.

Another quantity related to this aspect (via the localization of the 3d hole) is the spin-density on the central copper atom. 
In Table \ref{tab1}, we present the values of several LDA, GGA and hybrid DFT-derived Mulliken spin-densities (SD) on the central Cu atom 
and the {\it ab initio} CASSCF values for both states, each at its equilibrium geometry. Note that both CASSCF wavefunctions were optimized 
considering 21 active electrons (11 from Cu and 5 from each Cl atom) in 14 active orbitals, leading to large expansions with about 23~000 CSF.
Clearly, a rather different picture of the spin-density distribution is obtained with the DFT-derived methods {\it vs.} 
the corresponding {\it ab initio} ones, especially for the $^2\Pi_g$ state. 
It is quite remarkable that the quality of the excitation spectrum obtained 
with these functionals can be related to the magnitude of the spin-density on the central metal atom, since although all the 
functionals yield SD(Cu) values close to 1.0 for the $^2\Sigma_g^+$ state, the corresponding value for the $^2\Pi_g$ ground state 
shows large variations between the good and bad-performing functionals. 
The PBE0 SD(Cu) value is 0.64, while the BLYP and PBE96 spin-densities on copper are only 0.43. An intermediate situation arises 
for the next two best performing functionals, B3LYP and B97-2, with larger values of 0.57 and 0.54. The CASSCF(21,14) spin-densities 
are both very close to 1 for both electronic states, and this is precisely why it is generally thought that these ligand states actually 
present a much more localized hole on the central copper atom than any of the DFT descriptions provide. 
We shall address this important point in more detail in what follows.
\section{Ground-state spin density with DFT}
\label{dftrho}
\begin{figure}
\includegraphics[width=\columnwidth]{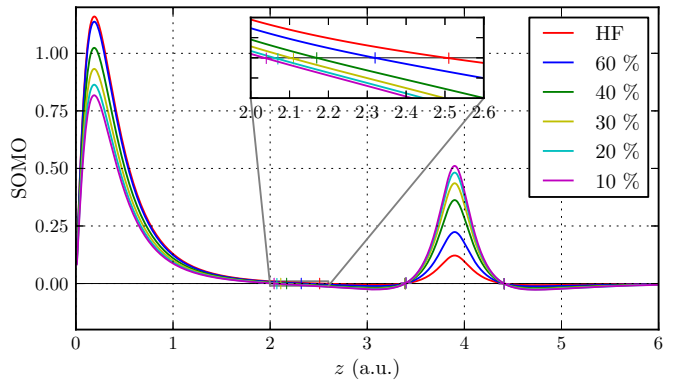}
\caption{Plot of the singly occupied molecular orbital along the nuclear axis $z > 0$ (copper at origin and chlorine at $z$=3.9 a.u.)
for the $^2\Pi_g$ ground-state using B3LYP as a function of the HF exchange percentage used in the hybrid functional. 
The values of $x$ and $y$ are fixed to 0. and 0.15, respectively. The inset is a blow-up of the region in the middle of the bond where the SOMO vanishes. 
See discussion in Sec.\ref{qmcrho}.}
\label{fig2}
\end{figure}

\begin{figure}
\includegraphics[width=\columnwidth]{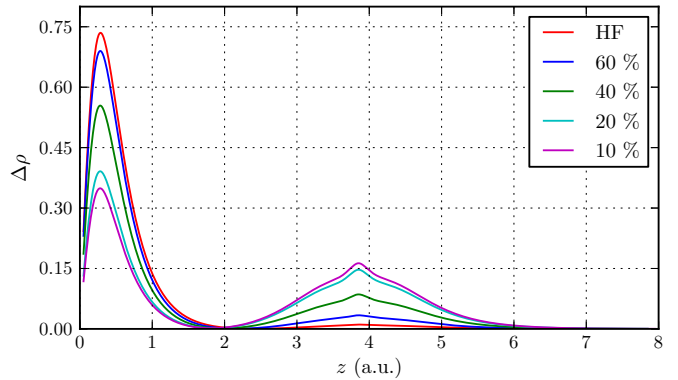}
\caption{Ground-state spin density with B3LYP as a function of the HF exchange percentage. Cu at the origin and Cl atoms located at z=$\pm$ 3.9 a.u.; only the positive z axis is shown}.
\label{fig3}
\end{figure}

Having in mind the previous results, we start our analysis of the ground state spin density along the molecular axis with the DFT approaches.
Note that the essential of chemical/physical properties of the molecule takes place via the singly-occupied molecular orbital (SOMO). 
It is so since in the spin-restricted Kohn-Sham formalism the contribution to the spin density resulting from all lower-lying orbitals cancels out and 
the local spin density is directly written as the square of the singly occupied orbital (other orbitals also contribute but in an indirect way through 
the Kohn-Sham optimization). The $\sigma$ or $\pi$ symmetry of the SOMO defines the overall symmetry of the ground state.
In Figure \ref{fig2} the SOMO orbital obtained with B3LYP for the $^2\Pi_g$ ground state as a function of the internuclear axis $z$
for different values of HF exchange percentage are shown, along with the ROHF orbital (100\% of HF exchange). 
The basis set used is that optimized in a previous work.\cite{RamDau-JCP04}
Since these orbitals are centro-antisymmetric 
with respect to the central Cu atom, only the $z>0$-region is shown. In this figure, the  $y$ coordinate is fixed to zero and $x$ to 0.15, a value close 
to the maximum of the highest peak of the orbital. Recall that in the present case of a one-determinant restricted KS or HF representation, 
the spin density as defined in Eq.\ref{deltarho} reduces to the integrated value over $(x,y)$ of 
the square of this latter orbital. Figure \ref{fig3} gives the B3LYP spin-densities obtained as a function of the HF exchange percentage 
together with the HF spin density.
The two-peak structure of the SOMO (Figure \ref{fig2}) and SD (Figure \ref{fig3}) is clearly seen, one peak localized very close
to the central Cu atom and the other on the chlorine atom. The relative height between the two maxima is strongly dependent on 
the percentage of HF exchange considered. In the case of HF (100\% of HF exchange) the main peak is the highest one while the 
secondary peak on Cl is 20 times smaller. This indicates a highly localized character for the 3d hole at the HF level. When decreasing the
percentage of HF exchange in B3LYP, the level of localization is found to decrease uniformly. Note also that the location of the zero (node)
of the SOMO in the middle of the bond (see blow-up in Figure \ref{fig2}) is very dependent on the level of HF exchange. This result 
will be discussed in the context of the FN-DMC results (see Sec.\ref{qmcrho}).
As in the case of the $^2\Sigma_g$-$^2\Pi_g$ transition energy presented above, there is no physically meaningful reason to 
decide which amount of non-local HF exchange should be used in B3LYP for this metallic system.

\section{Ground-state spin density with WFT: HF and beyond}
\label{wftrho}
\begin{figure}
\includegraphics[width=\columnwidth]{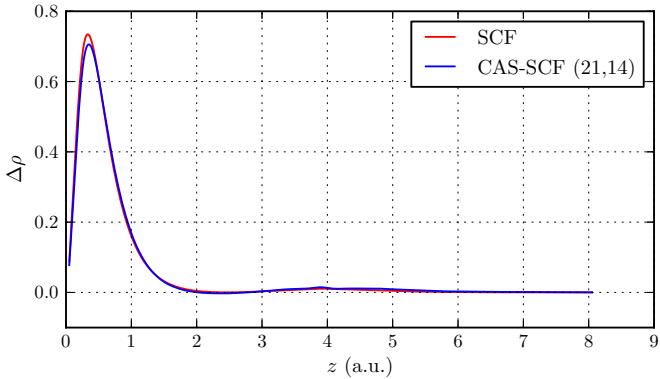}
\caption{
$^2\Pi_g$ state spin density along the molecular axis (in a.u.) at the HF-SCF and CASSCF levels.}
\label{fig4}
\end{figure}

Let us now turn our attention to the spin densities obtained from wavefunction approaches (WFT) at different levels of theory.
Figure \ref{fig4} shows $\Delta \rho(z)$ obtained for the $^2\Pi_g$ ground-state using both ROHF and CASSCF calculations.
In the later case the active space chosen includes 14 orbitals (3s and 3p shells of both Cl, 4s and 3d shells for Cu)
and 21 valence electrons (5 from each Cl and 11 from Cu) are distributed among them. In the resulting CASSCF(21,14) expansion 
the HF coefficient is found to be rather large (0.95), thus indicating a strong single-reference character of the ground-state wavefunction.
Therefore, dynamic correlation effects largely dominate this problem. Note that the CASSCF spin-density distribution 
presented has been obtained by considering only the first one hundred determinants corresponding to the largest coefficients in the 
expansion. As seen in Figure \ref{fig4} and expected from the single-reference nature of the wavefunction, HF and CASSCF spin-densities are 
almost identical.
In both cases the 3d hole is found to be strongly localized on the copper atom and almost no SD is present on the chlorine atoms.
As shown in Table \ref{tab1}, only the use of highly correlated methods (CCSD(T), CPF or ACPF) can recover the correct energetic ordering 
of the two lowest electronic states. Unfortunately, given the huge number of CSF (ca. $7\times 10^9$) considered in these approaches, 
the spin density distributions at these levels of theory are not available. 

\section{Ground-state spin density with quantum Monte Carlo}
\label{qmcrho}
In this section we report all-electron quantum Monte Carlo (QMC) calculations of spin-densities. Several versions of QMC have been introduced 
in the literature; however, they all rely on the very same ideas and differ only by technicalities. Here we employ a variant of the
Fixed-Node Diffusion Monte Carlo (FN-DMC) method defined with a constant number of walkers. 
For details the interested reader is referred to the original work.\cite{srmc}
In FN-DMC we are faced with two main sources of error: the statistical error inherent 
to any Monte Carlo approach and the fixed-node error. Other sources of errors are also present but they can be easily controlled 
and made negligible (see, \cite{encycloped}). By increasing the number $N$ of Monte Carlo steps
the $1/\sqrt{N}$-statistical error can be decreased as much as desired, at least in principle. 
In each application presented below, this error has been reduced to a level sufficient for our purposes. 
In contrast, the fixed-node approximation is much more challenging and its control 
is a crucial issue of present-day QMC approaches. It is known that the magnitude of the fixed-node error 
is directly related to the quality of the nodal structure of the approximate trial wavefunction used 
in the simulation [the nodes are the $(3N-1)$-dimensional zeroes of the $3N$-dimensional wavefunction, $N$ being the number of electrons].
Exact total energies can be obtained only when using trial wavefunctions having the nodes of the exact (unknown) wavefunction.
It should be emphasized that, in contrast with the statistical error which can be reduced as desired by increasing Monte Carlo statistics, 
the fixed-node error is a {\it systematic error} (i.e., a bias) that survives even for infinite statistics.
Numerical experience has shown that, although fixed-node energies are very accurate, 
non-negligible errors on energy differences may still occur due to improper cancellation of fixed-node errors. 
Unfortunately, in some cases this error can be large enough to lead to {\it qualitative} wrong conclusions.
When considering closed-shell systems with a strong single-reference nature, nodal hypersurfaces
resulting from single-determinant representations ({\it e.g.}, Hartree-Fock or Kohn-Sham type) are expected 
to be of sufficient quality. As we shall see, in the CuCl$_2$ case considered here, the situation is different. Although the exact wavefunction
has a strong single-reference character, the presence of an open-shell makes the nodal structure of the wavefunction more 
difficult to describe. In this case the nodes of the $3N$-dimensional wavefunction turn out to 
be very sensitive to the 3-dimensional nodal pattern chosen for the singly-occupied molecular orbital (SOMO).
Note that it is an interesting case where the highly-complicated $3N$-dimensional nodes usually so difficult to visualize 
can be reduced, in a good first approximation, to a much simpler 3D-pattern.

A last point to specify is the way spin-densities are computed here. In the case of total energies 
it is known that the only systematic error is the fixed-node one, despite the fact that the stationary diffusion Monte Carlo distribution 
is not exact. DMC actually samples the so-called mixed distribution given by the product of the trial wavefunction and 
the exact wavefunction, see \cite{encycloped}.
In the case of properties other than energies, this is no longer true and some additional error related to 
the trial wavefunction contribution in the mixed distribution is present.
This error can be removed in different (costly) ways, see {\it e.g.} \cite{caf88} and \cite{rep-mc}. However, 
such a possibility was not considered here since, as we shall see later, the dominant source of error is the fixed-node approximation. 
Spin-densities are thus calculated in a standard way using a hybrid second-order estimate. For a general observable $O$
it is expressed as\cite{lestbook}
\begin{equation}
\langle O \rangle \sim  2 \langle O\rangle_{\rm DMC} - \langle O \rangle_{\rm VMC}
\label{obs}
\end{equation}
where averages are taken either over the mixed distribution sampled in DMC or over the 
squared trial wavefunction density sampled in a variational Monte Carlo (VMC) simulation.
Here, the properties to be computed are the $\alpha$ and $\beta$ spin densities and the quantities to average are merely 
the number of $\alpha$ or $\beta$ electrons falling within histogram bins.

\begin{figure}
\includegraphics[width=\columnwidth]{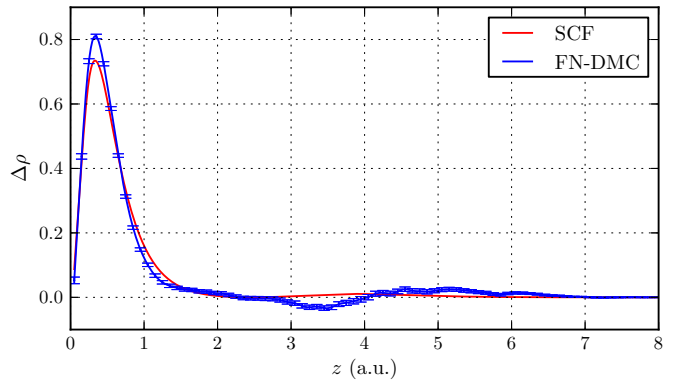}
\caption{
$^2\Pi_g$ state spin density along the molecular axis (in a.u.) at the FN-DMC level with a HF trial wavefunction}
\label{fig5}
\end{figure}

\begin{figure}
\includegraphics[width=\columnwidth]{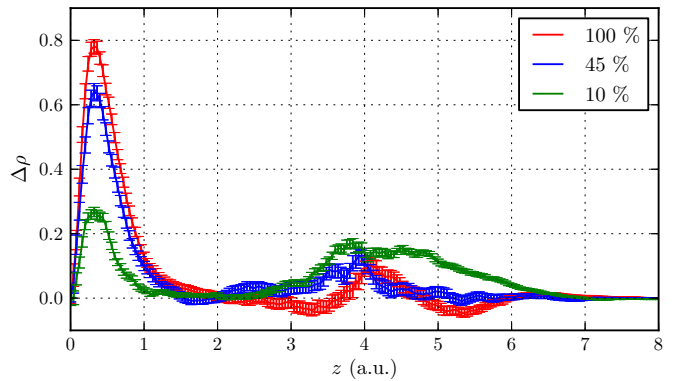}
\caption{$^2\Pi_g$ state FN-DMC spin densities along the molecular axis (in a.u.) using as trial wavefunction B3LYP-KS determinants 
obtained with different amounts of HF exchange}
\label{fig6}
\end{figure}

\begin{figure}
\includegraphics[width=\columnwidth]{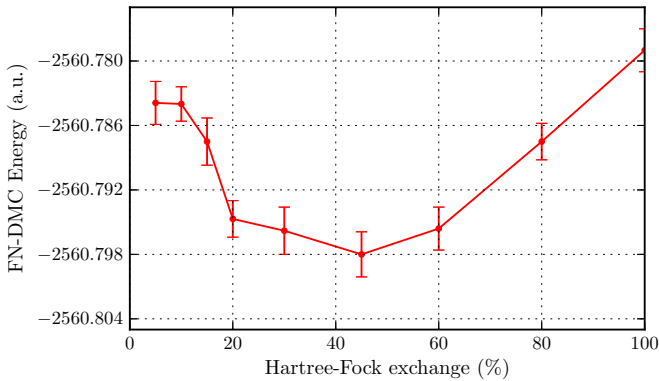}
\caption{Total FN-DMC $^2\Pi_g$ ground state energy obtained with a B3LYP determinant as a function of the HF exchange percentage for $CuCl_2$.}
\label{fig7}
\end{figure}

The all-electron Fixed-Node DMC spin-density for the $^2\Pi_g$ ground-state using a 
Hartree-Fock wavefunction as trial wavefunction (complemented with a standard Jastrow factor to reduce statistical fluctuations for the energy)
is presented in Figure \ref{fig5}. The basis set used is that optimized in a previous work.\cite{RamDau-JCP04}
For comparison, the Hartree-Fock SD is also given.
Although some differences between the two curves exist, they should essentially be considered as the same when compared to 
the typical differences present in the DFT-SD curves, Figure \ref{fig2}. The small differences include a slight increase of the main DMC peak
and a small ``spin-density wave'' around Cl atoms. In this calculation the total energies obtained at the SCF and FN-DMC levels are -2558.1050 and 
-2560.719(2), respectively. To get an assessment of the accuracy reached here with QMC a rough
estimate of the exact total energy of the molecule can be done. For that we add to the sum of atomic energies the atomization energy 
calculated at the SCF level. Taking for Cl the value from Davidson {\it et al.},\cite{davidson} for Cu the HF energy of Bunge,\cite{bungescf}
plus the correlation energy estimate of Clementi {\it et al.}\cite{clementi}, the exact energy of separate atoms is found 
to be about -2560.868. Adding the SCF atomization energy we get a total ground-state energy for CuCl$_2$ of about -2561.045 a.u.. The percentage
of correlation energy recovered by FN-DMC with HF nodes is thus quite large, roughly $\sim$ 89\%.
Thus, with this highly-correlated description of the wavefunction but imposing HF nodes, it is found that the shape of the SD
is not quantitatively changed with respect to that obtained at the SCF level.
Let us now consider the FN-DMC spin-densities obtained when using KS determinants instead of the Hartree-Fock one as trial wavefunctions.
The KS determinants were obtained with standard B3LYP and with B3LYP with a variable amount of HF exchange. 
In Figure \ref{fig6} the corresponding FN-DMC spin-densities are presented.
As clearly seen, the overall shapes of FN-DMC spin-densities are tightly correlated with those obtained at the corresponding DFT level,
see Figure \ref{fig3}. The results are thus similar to what has just been obtained for the SCF case: No qualitative change of the spin densities 
is obtained when passing from the variational to the FN-DMC level. These results strongly suggest that the key factor
determining the SD shape is the nodal structure of the trial wavefunction used. The situation can thus be summarized as follows: 
i.) The amount of Hartree-Fock exchange in B3LYP determines the relative weight of 3p$_{\rm Cl}$ and 3d$_{\rm Cu}$ atomic contributions to the
SOMO ii.) the nodes of the SOMO are directly related to this relative weight iii.) the nodal pattern of the whole trial wavefunction 
is dominated by the SOMO nodes. In the inset of Figure \ref{fig2} a blow-up of the SOMO in the region around 
its node located at the middle of the Cu-Cl bound is presented; the two 
other nodes close to the secondary peak are weakly dependent on the level of exchange and will not be discussed here.
The position of the central node is seen to be very sensitive to the percentage of HF exchange. Its location ranges 
from $R_{\rm node}= 2.5$ for the Hartree-Fock wavefunction to about 2.05 for the KS determinant corresponding to the lowest HF percentage of 10\%.
In short, the nodes of the trial wavefunction are very sensitive and directly related the amount of HF exchange chosen.
At this point, the situation is clearly not satisfactory since the overall shape of spin distributions 
is determined by the specific choice of nodes of the SOMO. Said differently, FN-DMC is not able to change qualitatively the global 
features of the spin-density associated with the approximate trial wavefunction given in input for the diffusion Monte Carlo process. 
We thus need to resort to alternative approaches capable of changing the nodes when electronic correlation effects are included. 
This will be the subject of the following section. Before that, let us nevertheless note that there exists in FN-DMC an internal criterion
for estimating the nodal quality. It is based on the variational principle stating that the ``better'' the nodes are, the lower the fixed-node 
energies are expected to be.\cite{lestbook} Figure \ref{fig7} presents the variation of the total FN-DMC ground-state energy as a function of 
the amount of exchange considered. The basis set employed in these calculations is that of Weigend and Ahlrichs,\cite{weigend} 
which leads to significantly lower fixed-node energies than those obtained with the basis set proposed in Refs.\cite{RamDau-JCP04} and \cite{RamDau-JCP05}.
Quite remarkably, a minimum is observed for a HF exchange percentage around $45\%$. 
This result is interesting and may be understood as a first indication of the typical amount of HF exchange that should be employed. However,
let us stress that this result must be considered with lot of caution since the sensitivity of the FN-DMC results on nodal choice
is high and optimizing only the one-dimensional nodes of the SOMO could be insufficient. Furthermore, optimizing nodes via minimization of 
the total energy is not a guarantee of improvement for other properties like spin density distributions.

\section{Ground-state spin density with near-FCI}
\label{cirho}
In this section we report near-Full Configuration Interaction (FCI) calculations for total energies and spin-densities. 
To achieve converged results on this system including 63 electrons, a small 6-31G basis set is employed for both atoms, 
leading to a molecular basis of 51 orbitals. Although the accuracy reached with such a modest basis 
set can be questioned, it will allow us to investigate the major trends obtained when using multi-determinantal wavefunctions 
whose nodal structure may change when correlation effects are introduced which, as we have shown, is not possible to do with FN-DMC. 
To realize FCI-type calculations for this system we use the CIPSI approach (Configuration Interaction with Perturbative Selection done Iteratively), 
a method proposed more than four decades ago (see, \cite{malrieu},\cite{evan83}, and more references in \cite{canada}) 
and very recently introduced in the context of QMC approaches.\cite{canada} 
For a detailed presentation of this approach the reader is referred to the original works. 
In short, CIPSI is a variational and multireference perturbational configuration interaction approach in which determinants that are to be 
included in the variational space are selected iteratively according to an energy criterion. 
Determinants perturbationally generated are added to the variational wavefunction when their perturbative contribution to the total 
energy is greater than a given threshold. In contrast with standard CI approaches where a whole set of particle-hole excitations are considered 
(single-excitations, single- and double-excitations, etc.), only excitations having a significant impact on the wavefunction expansion 
are selected as variational contributions. The relevance of a particular excitation is decided by comparing its energy contribution 
with the pre-fixed threshold. This procedure is applied iteratively until a given target number of determinants is reached.
In practice, this leads to rather compact variational expansions consisting of a limited number of 
determinants in each type of excitations. Furthermore, higher-degree excitations not usually present in standard CI expansions may also 
be naturally introduced in the variational space with the CIPSI approach. Finally, let us note that several applications for a variety of 
metal-containing molecules have been realized during the 90's, see {\it e.g.} \cite{cux},\cite{agx},\cite{ag3}. 
The major difference between these applications and the present study is the size of the variational space that is taken much 
larger here (up to a million of determinants). The second-order perturbational correction is thus much reduced and an accuracy close 
to the FCI limit can be reached in the present application.
\begin{figure}
\includegraphics[width=\columnwidth]{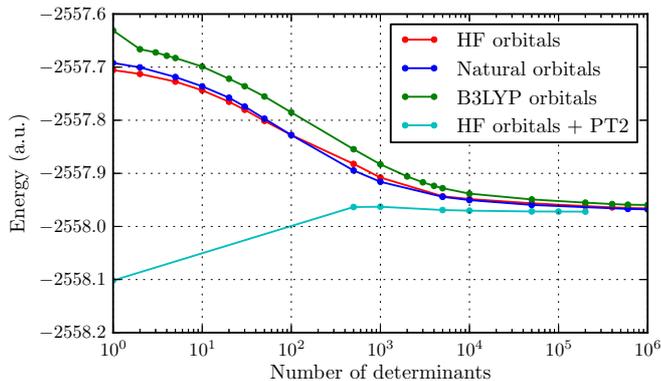}
\caption{Total variational (three upper curves) and variational + perturbational (lower curve) ground-state energy as a function of the 
number of determinants kept in the CIPSI selection process. SCF, B3LYP and natural orbitals are used.}
\label{fig8}
\end{figure}
In Figure \ref{fig8} the convergence of the ground-state energy as a function of the number of determinants kept in the variational space
is presented. To reduce the size of the variational CI calculation, molecular orbitals of the neon and argon cores for the chlorine 
and copper atoms, respectively, have been kept frozen. Calculations have been performed using 36 active molecular orbitals and 
25 valence electrons. We stress that the size of the full CI space is about 
$10^{18}$ determinants. 
With the present basis set the maximum number of determinants in the variational space considered here is $10^{6}$.
The three upper curves of Figure \ref{fig8} are the variational energy curves corresponding to the multi-determinantal expansion,$|\Psi_0\rangle$ 
built using either SCF, DFT-B3LYP, or natural molecular orbitals. The latter were constructed from the variational CIPSI wavefunction 
obtained with $10^{6}$ determinants. As seen on the figure, all curves are found to converge almost to the same value, as it should be 
when approaching the full CI limit. The lower curve shows the so-called CIPSI energy obtained by adding to the variational energy $E_0$ 
the second-order perturbative contribution defined as
\begin{equation}
{\rm E}_{PT2}=-\sum_{i \in P} \frac{  { {\langle \Psi_0|H|D_i} \rangle}^2}{\langle D_i|H|D_i\rangle - E_0}.
\label{ept2}
\end{equation}
where $P$ denotes the set of all determinants not present in the multi-determinantal expansion $|\Psi_0\rangle$ but connected
to it by the Hamiltonian $H$ (single- and double-excitations). For clarity only the CIPSI curve obtained with HF orbitals
is shown, the other CIPSI curves have a similar behavior. ${\rm E}_{PT2}$ can be considered as a measure of the energy difference
between the variational energy and FCI limit. As seen on the figure, the convergence of CIPSI energy is particularly rapid, 
we consider the limit has been attained with about 20~000-50~000 determinants in the variational space. 
For a large enough number of determinants, the perturbative correction ${\rm E}_{PT2}$ becomes quite small,
this being a reliable indicator of the convergence to the FCI limit.
\begin{figure}
\includegraphics[width=\columnwidth]{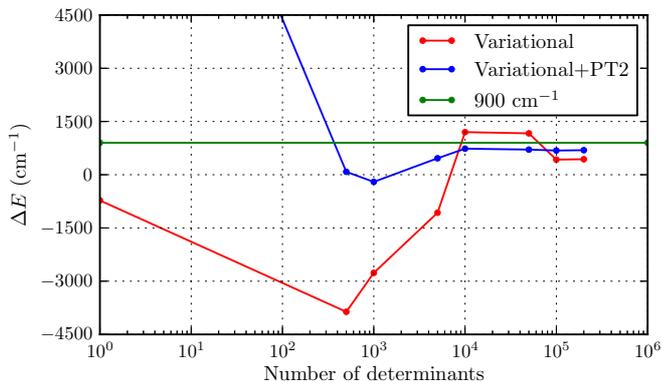}
\caption{$^2\Sigma_g-^2\Pi_g$ energy gap as function of the number of determinants in the variational space.}
\label{fig9}
\end{figure}
In Figure \ref{fig9} the energy difference between the $^2\Sigma_g$ and $^2\Pi_g$ states as a function of the number of determinants is
presented (note the logarithmic scale for the number of determinants). The evolution of the energy difference is shown both for
the variational energies and the CIPSI energies (variational + ${\rm E}_{PT2}$).
At the purely variational level the $^2\Sigma_g$-$^2\Pi_g$ gap starts with a negative value (as it should be for one-determinant 
SCF wavefunctions, see Table \ref{tab1}) and then changes sign when about 6~000 determinants are variationally included. 
With a larger number of determinants the energy gap is found to converge to a 
value close to about 1~000~cm$^{-1}$. At the CIPSI level, the convergence is even faster and better behaved. The CIPSI limit is
close to the variational one. Note that for a small number of determinants the second-order energy correction is large and unphysical. 
CIPSI results are only meaningful in the large number of determinants regime, where the second-order contribution is indeed a correction. 

Results obtained for the energy are very satisfactory; they demonstrate 
that nearly-FCI calculations are able to describe the transition between the two lowest electronic states despite the smallness of the basis set, 
since the converged value obtained for the energy gap is very close to the estimated SO-deperturbed value of about 900~cm$^{-1}$.
\begin{figure}
\includegraphics[width=\columnwidth]{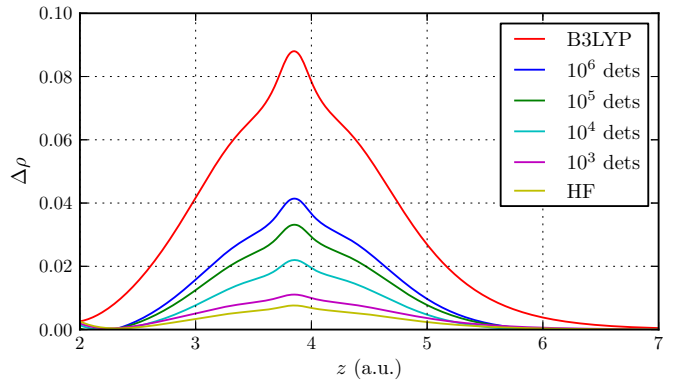}
\caption{
$^2\Pi_g$ state. Spin density around the chlorine atom with HF, B3LYP and CIPSI with varying number of determinants. The basis set employed 
is 6-31G.}
\label{fig10}
\end{figure}
For a deeper analysis we have also calculated the spin-density obtained from the CIPSI variational wavefunction. Its evolution
as a function of the number of selected determinants is plotted in Figure \ref{fig10}. 
As usual, a two-peak structure is observed. In the figure only data for the secondary peak on the chlorine atoms are shown. 
For comparison, the SD obtained at the SCF (nearly vanishing small peak) and B3LYP levels (the highest peak) are also plotted. 
Remark that the maximum of the B3LYP peak is about 0.087, to compare with the value of 0.15 for the very same quantity 
presented in Figure \ref{fig3} (ordinary B3LYP with 20\% of HF exchange). We have checked that this difference results from  basis set 
incompleteness due to the limitations of the 6-31G basis sets on both atoms. The CIPSI spin densities lie between both extreme curves and 
the height of the SD peak is found to increase continuously with the number of determinants in the variational space. 
For the maximum number of determinants of $10^{6}$ the height of the peak
is not yet fully converged but is large and represents about 40\% of the B3LYP peak.
\begin{figure}
\includegraphics[width=\columnwidth]{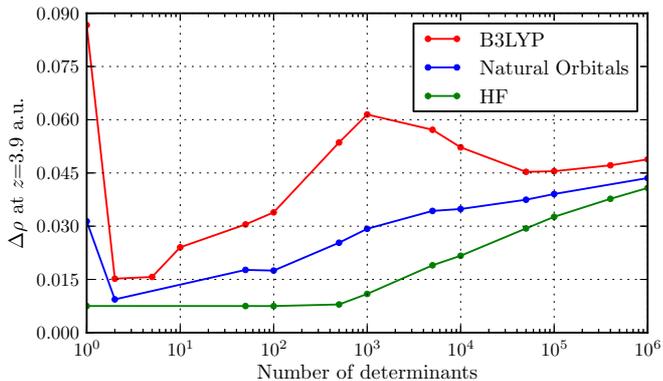}
\caption{$^2\Pi_g$ state. Convergence of the maximum of the secondary peak of the spin distribution as a function of the number of determinants in the variational space. Different types of molecular orbitals were used.}
\label{fig11}
\end{figure}
In Figure \ref{fig11} we present a more complete view of the convergence of the secondary peak as a function of the number of determinants selected, 
and with various types of molecular orbitals used in CIPSI. When using HF molecular orbitals, the spin density is found to 
remain close to zero up to a thousand of determinants and then begins to increase uniformly until it attains its maximum value. 
With B3LYP molecular orbitals the situation is qualitatively different. Starting from one determinant from a high value of the peak 
(as shown above with the B3LYP KS determinant), it decreases rapidly to a value close to zero. This phenomenon can easily be interpreted 
by noting that in a CI calculation the role of the first determinants consists essentially in lowering the energy via single-excitations 
whose effect is to optimize in an effective way the (natural) one-body orbitals (here, going from pure KS to SCF-type orbitals). 
Because of that, the SD is first found to almost vanish like in a SCF calculation. Next, when more determinants are added to the variational wavefunction, 
dynamical correlation contributions begin to enter the game (typically, through two-particle excitations) and then the SD starts to increase. 
Directly using natural orbitals, a similar phenomenon occurs but in a less marked way since 
the initial SD value is smaller than in the DFT case.
\begin{figure}
\includegraphics[width=\columnwidth]{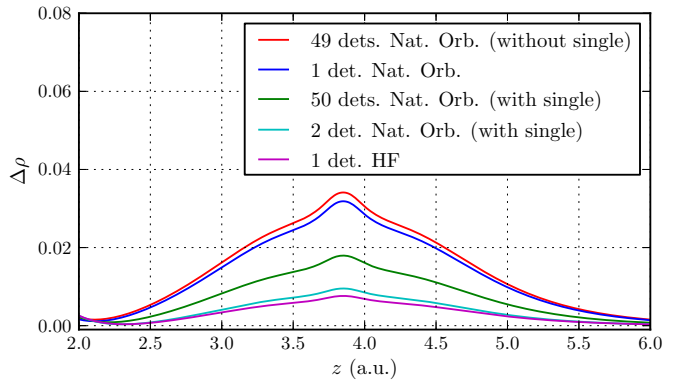}
\caption{Comparison of the secondary peak of the SD obtained with a small number of determinants, including or excluding the SOMO-LUMO single-excitation.}
\label{fig12}
\end{figure}
To support the previous scenario regarding the role of single-excitations, we present in Figure \ref{fig12} the shape of the secondary 
peak for a small number of determinants (about 50 determinants) using natural orbitals and including or excluding the SOMO-LUMO single-excitation 
that enters first in the variational space. Using only
one determinant built with natural orbitals the maximum found for the peak is about 0.035, the largest value of the figure. Now, a short CIPSI 
calculation including only 50 determinants in the variational space is performed. Two situations are considered depending on the fact 
that the determinant representing the SOMO-LUMO single-excitation is removed or not from the variational expansion. In the first case the peak 
is essentially unchanged. In sharp contrast, in the second case the peak of the spin distribution is significantly reduced 
and many additional determinants are needed to recover its original shape. These results nicely illustrate the role of 
single-excitations recovering the Hartree-Fock nature of the orbitals when a small number of determinants is considered in the reference space.

To conclude, results for the spin-density distribution along the molecular axis obtained using near-FCI calculations with the small 6-31G basis set lead to a particularly clear picture.
At the SCF (or CASSCF) level, no spin density is present on the chlorine atoms. The Cu(3d) hole is strongly localized on the copper atom and 
the ground-state is of symmetry $^2\Sigma_g$ (i.e., a negative energy gap with our definition). However, when introducing dynamical electron 
correlation contributions via a nearly full-CI calculation, the situation changes dramatically. The $^2\Pi_g$ becomes the ground-state 
and the Cu(3d) hole is found to be partly delocalized over the chlorine atoms. Quantitatively, the SD peak located on the Cl atoms is roughly 
two times smaller than the peak obtained with standard B3LYP (with 20\% HF exchange contribution) and is similar to that
obtained using B3LYP but with a higher percentage of HF exchange of about 40\% (see, Figure \ref{fig3}). 
A remarkable point is that this result appears to corroborate the 
fact that in FN-DMC optimal nodes were obtained with KS orbitals built using about the same percentage of HF exchange
(see, Figure \ref{fig7}). Finally, we have made some attempts to extend these near-full CI calculations to larger basis sets (triple-$\zeta$+polarization). 
Unfortunately, given the much larger size of the full-CI space, we were not able to reach a level of convergence sufficient to get reliable information.
\section{Discussion and summary}
\label{discussion}
In this work calculations of the total energy and spin-density for the $^2\Pi_g$ ground state of the CuCl$_2$ molecule have been presented
using various quantum-mechanical methods. Depending on the approach employed different qualitative and quantitative descriptions of the 
spatial distribution of the spin density along the molecular axis have been found. At the root of such discrepancies lie the different ways the 
electronic structure is described and approximated.

At the DFT level the description of the low-lying states of the molecule is very dependent on the type of exchange-correlation functional chosen.
Recalling that hybrid functionals had been shown to provide the best agreement with experimental data, the energy gap between the 
two lowest states, $^2\Pi_g$ and $^2\Sigma_g$, is found to be very sensitive on the fraction of HF exchange used in the functional.
For B3LYP the gap is roughly linearly dependent on this fraction, starting from a $^2\Sigma_g$ ground-state with 100\% HF exchange 
to a $^2\Pi_g$ ground state with a sufficiently low percentage (about 40\% and less). Regarding the spatial distribution of the spin density 
a similar strong dependence on the fraction of HF exchange used in the hybrid functional is found.
In the case of the CuCl$_2$ molecule having a single unpaired electron, the DFT spin density is entirely determined by the square of the SOMO orbital.
By varying the HF exchange percentage, the shape of this orbital may be continuously varied and so is the spin density. 
With the full HF exchange, the DFT spin-density is almost entirely localized on copper, while lower levels of HF exchange lead to increasingly 
delocalized spin densities on both Cl ligands. Such results are clearly disturbing, since there exists no internal criterion within DFT 
to decide which amount of HF exchange should be used and, thus, a meaningful chemical picture of the electronic distribution is difficult to obtain.
We recall that in the DFT framework the self-interaction error (SIE) is known to be directly related to the exchange part of the functional; 
in the case of the metal-containing molecules with a high electronic density in the $d$ shell this error may be particularly important 
and not easy to control, thus leading to a potentially incorrect description of the delocalization of electronic distributions.

A common way of shedding light on a situation where DFT leads to unpredictable results is to resort to highly-correlated 
post-Hartree-Fock methods where the construction of accurate {\it 3N}-dimensional wavefunctions allows, in principle, a better control 
of the details of the electronic structure but at a much higher computational price. At the HF level, and in agreement with hybrid DFT results 
with full HF exchange, the spin density is found to be completely localized on the central copper atom. At the CASSCF level including
all Cl(3p) and Cu(3d,4s,4p) orbitals as active orbitals, the wavefunction is not significantly changed and is largely dominated by the HF 
determinant. In other words, the dynamical correlation effects are dominant here and the spin density calculated with CASSCF is practically
identical to that of the HF description. Unfortunately, as illustrated by a number of works, it is very difficult to reproduce with sufficient 
accuracy the dynamical correlation effects  and thus to give a quantitative description of the low-lying states; in particular, to obtain 
the correct energy difference between the two lower states requires very high level calculations [{\it e.g.} CCSD(T) or ACPF] with large optimized basis 
sets. Unfortunately, these methods do not provide the final electronic density that would allow us to conclude 
on the true chemical picture concerning the spin density. 

To escape from such limitations, we have proposed to resort to QMC calculations that are known to be particularly accurate. Using 
different types of trial wavefunctions fixed-node DMC calculations 
of both ground-state total energies and spatial distributions of the spin density have been performed. Unfortunately, 
although we get state-of-the-art total energies (we have estimated that about 90\% of the total correlation energy for the ground-state energy is recovered),
spin densities calculated within the fixed-node approximation are found to be too dependent on the nodal structure of the trial 
wavefunction employed. To be more precise, in the present case with a singly occupied molecular orbital (SOMO), we have seen that the 
complex {\it 3N}-dimensional nodal hypersurface of the full trial wavefunction is dominated by the 3-dimensional nodes of the SOMO 
and that the shape of the FN-DMC spin densities calculated is directly related to the shape of this orbital. 
Stated differently, qualitatively different spin density distributions can be obtained even at the supposedly very accurate 
FN-DMC level, depending on the choice of the singly-occupied orbital used in the trial wavefunction. 
Using a HF-type wavefunction, the FN-DMC spin density 
closely resembles to that obtained at the variational HF level. Similarly, when using various SOMO KS orbitals obtained with a variable exchange 
hybrid DFT method (here, B3LYP), FN-DMC spin densities resembling to their KS counterparts are obtained. As a consequence, 
it becomes impossible to decide on such grounds what is the correct chemical picture for the spin distribution. Nevertheless, 
we have noted that, within the framework of FN-DMC approaches, there exists an internal criterion allowing to estimate the nodal quality: 
The lower the fixed-node energy is, the ``better'' the nodes  are expected to be (variational property of the fixed-node energy, see \cite{lestbook}). 
We have insisted on the fact that this criterion should be taken with lot of caution for a property other than the energy; 
however, it is worth noting that the nodes of the SOMO minimizing the fixed-node energy are those corresponding to a contribution 
of HF exchange of about 40-45$\%$, considerably higher than the ordinary B3LYP but much smaller than pure HF.

In order to elucidate these various contradictory results and to get a plausible description of the ground-state spin-density distribution,
we have proposed to perform near-full Configuration Interaction calculations. Only such calculations can indeed yield a reliable 
balance between electron correlation and exchange effects. Obviously, in the present case where the molecule contains 63 electrons,
ordinary FCI calculations using standard basis sets are just unfeasible. However, to circumvent this difficulty we have proposed to greatly reduce
the dimensionality of the CI problem by employing the small 6-31G basis sets on both atoms and then by performing selected multireference CI
calculations, which allowed us to avoid huge intractable multi-determinantal variational FCI expansions. Clearly, by using a small 
basis set the quantitative accuracy of the results can be questioned. However, since all types of electronic excitations are considered\cite{note}, 
it can be expected that chemically relevant trends regarding the various aspects obtained above with other methods may emerge.
The multireference CI used here is a perturbatively selected CI scheme (CIPSI) that includes in a hierarchical way the most important 
determinants to asymptotically approach the FCI limit. Remarkably, it has been shown that using CIPSI for both the ground and first-excited state, 
these near-FCI calculations yield transition energies and spin-densities that are almost converged.

From the all set of data obtained for total energies, energy gap, spin densities, and the dependence of the various results on the number 
of determinants and types of molecular orbitals used, a rather coherent chemical picture emerges. 
At the uncorrelated (SCF) level, the lowest state is of $^2\Sigma_g^+$ symmetry and the Cu(3d) hole is completely localized on the copper atom. 
When dynamical correlation effects are added the ordering between the $^2\Sigma_g$ and $^2\Pi_g$ states is reversed and the hole is found to 
partly delocalize over the Cl ligands. At the ordinary DFT-B3LYP level the Cu(3d) hole is too delocalized over the chlorine atoms due 
to an improper balance between the self-interaction and exchange effects. To get a chemically meaningful description of electronic distributions 
using B3LYP-DFT the percentage of HF exchange used must be increased up to about 40\%. At the fixed-node DMC level, spin densities are 
found to be intimately related to the shape of the singly occupied molecular orbital, an orbital whose nodes are in turn directly 
related to the level of HF exchange used to derive it. Using as criterion the minimization of the FN-DMC ground-state energy, the optimal nodes for the SOMO 
are obtained for a HF exchange weight of about 40\%, a result coherent with what has been obtained with near-FCI. Finally, let us note that 
the fact that DFT overestimates delocalization effects of magnetic holes in molecular systems has already been noticed in the literature by other authors 
(see, {\it e.g.} \cite{malrieu2}).\\
\\
{\it Acknowledgments.}
We would like to acknowledge a fruitful discussion with Jeremy Harvey about the role of the exchange in metal-containing molecular systems. 
The Agence Nationale pour la Recherche (ANR) is thanked for support through Grant No ANR 2011 BS08 004 01. 
ARS thanks support from the 2013 CONACYT (Mexico) sabbatical program and from the Professeur Invit\'e program 
from the Universit\'e de Toulouse (UPS). This work has been made possible through generous computational support from CALMIP
(Toulouse) under the allocation 2012-0510, GENCI, and CCRT (CEA).


\begin{thebibliography}{99}
\bibitem{note}
In practice, excitations up to quintuples are treated variationally, those up to septuples perturbatively, the others having a 
negligible contributions.
\end{thebibliography}
\end{document}